\documentclass[iop]{emulateapj}
\usepackage{bm,natbib,graphicx}
\newcommand{\mv}[1]{{\bm #1}}
\newcommand{\fobs}{\mathcal{F}_{\rm obs}}
\newcommand{\fint}{\mathcal{L}_{\rm int}}
\newcommand{\rmin}{R_{\rm min}}
\newcommand{\rmax}{R_{\rm max}}
\shorttitle{FRB population statistics}
\shortauthors{Vedantham et al.}
\begin{document}

\title{The Fluence and Distance Distributions of Fast Radio Bursts}
\author{Vedantham H.~K., Ravi V., Hallinan G.}
\affiliation{California Institute of Technology, 1200 E California Blvd., Pasadena CA 91125; harish@astro.caltech.edu}
\author{Shannon R.~M.}
\affiliation{CSIRO Astronomy and Space Science, Australia Telescope National Facility, PO Box 76, Epping, NSW 1710, Australia}

\begin{abstract}
Fast radio bursts (FRB) are millisecond-duration radio pulses with apparent extragalactic origins. All but two of the FRBs have been discovered using the Parkes dish which employs multiple beams formed by an array of feed horns on its focal plane. In this paper, we show that (i) the preponderance of multiple-beam detections, and (ii) the detection rates for varying dish diameters, can be used to infer the index $\alpha$ of the cumulative fluence distribution function (the log$N$-log$F$ function: $\alpha=1.5$ for a non-evolving population in a Euclidean universe). If all detected FRBs arise from a single progenitor population, multiple-beam FRB detection rates from the Parkes telescope yield the constraint $0.52<\alpha<1.0$ with $90$\% confidence. Searches at other facilities with different dish sizes refine the constraint to $0.66<\alpha<0.96$. Our results favor FRB searches with smaller dishes, because for  $\alpha<1$, the gain in field-of-view for a smaller dish is more important than the reduction in sensitivity. Further, our results suggest that 
(i) FRBs are not standard candles, and (ii) the distribution of distances to the detected FRBs is weighted towards larger distances. 
If FRBs are extragalactic, these results are consistent with a cosmological population, which would make FRBs excellent probes of the baryonic content and geometry of the Universe.
\end{abstract}
\maketitle

\section{Introduction}

Fast radio bursts (FRBs) are millisecond-duration, intense ($\sim 1$~Jy~ms) radio bursts that have dispersion measures (DMs) well in excess of expected Milky Way contributions \citep{lorimer2007, thornton2013, ravi2015, champ2015, petroff2015, spitler2014, masui2015a,keane2016}. Although the progenitors of FRBs and the associated emission mechanisms are interesting in their own right, the apparent extragalactic origin, if true, can be used to probe the intergalactic medium and the geometry of the Universe \citep{mcquinn2014, masui2015b, deng2014,zheng2014,kulkarni2014}. All reported bursts have been detected using single-dish telescopes that lack the angular resolution to obtain meaningful localizations, or even conclusively rule out a near-field or atmospheric origin. However (i) the strong adherence of FRBs to the dispersion and scattering laws expected from propagation through cold, turbulent plasma, and (ii) the measurement of Faraday rotation measure \citep{masui2015a} consistent with a magnetic field strength many orders of magnitude weaker than the 
terrestrial field, and (iii) detection of repeating bursts with a consistent sky poistion and dispersion measure \citep{spitler2016}, all favor an astrophysical origin. \\

An important attribute of any astrophysical population is the integral source counts, or the log$N$-log$F$ curve, which is the number of sources expected to have an observed fluence, $\fobs$, above a certain threshold. We model the log$N$-log$F$ curve as a power law with index $\alpha$: 
\begin{equation}
\label{eqn:lnls}
\mathcal{N}(>\fobs) \propto \fobs^{-\alpha}.
\end{equation}
where $\alpha=1.5$ for a non-evolving population in Euclidean space. 
We show that for far-field events, the fraction of events detected in multiple focal-plane feeds on a given telescope is  mostly determined by the index, $\alpha$. In particular, source counts with flatter slopes (values of $\alpha$ closer to zero) yield a relative abundance of brighter events which results in an increased propensity for multiple-beam detections. The principal motivation for this paper is a 
surprising large fraction (2 out of 16) of multiple-beam FRB detections with the Parkes multi-beam receiver. We use simulated far-field beam models of the Parkes multibeam receiver and the observed multiple-beam detection rates to constrain the value of $\alpha$ 
(Section~\ref{sec:alpha_mbs}). While doing so, we fully account for possible detections of bright events beyond the nominal beam Full Width at Half Maximum (FWHM), and also ensure that our results are robust to survey incompleteness at low fluence levels. \\

A flatter log$N$-log$F$ distribution that yields many bright events also makes telescope sensitivity less important than the field of view, i.e., a smaller dish may discover more FRBs than a larger one! Many authors have reported (non-)detections from telescopes with varying dish sizes \citep{siemion2012, spitler2014, saint2014, law2015} at L-band (around 1.4 GHz). The common frequency band used by these surveys allows us to compute their respective detection rates in a way that is largely independent of the inherent spectral or scattering properties of FRBs. We refine our Parkes multi-beam constraints with independent bounds on the value of $\alpha$ from such (non-)detections (Section~\ref{sec:alpha_mt}). \\

Finally, we discuss the implications of our bounds on $\alpha$ (Section~\ref{sec:disc}). The inconsistency we find of the FRB log$N$-log$F$ distribution with a non-evolving source population in Euclidean space has some important implications. To explore these implications, we consider a simple scenario where the FRB population cuts-off at some minimum and/or maximum distance. We find that FRB detection rates are either unbiased with or weighted towards larger distances to the progenitors. This bodes well for the use of FRBs as cosmological probes, even with telescopes with modest collecting areas.
\section{Multiple-beam detection statistics}
\label{sec:mbs}
\subsection{Beam-pattern calculations} 
We now describe the simulation set-up used to compute the beam patterns for a dish with multiple feed-horns in the focal plane. Our simulations do not assume a far-field geometry; the telescope-source distance is left as an input parameter. We have done so to facilitate future studies of terrestrial and atmospheric transients \citep{dolin2014,katz_perytons,danish2014}, which may be of great interest to atmospheric physicists, and at the very least, form a source of foreground `confusion' to the astronomer.\footnote{We point the interested reader to a intriguing study by \citet{close2010} of radio transients caused by meteor impact on spacecrafts.}\\

The multi-beam receivers on the Parkes \citep{smith1996} and Arecibo \citep{cordes2006} dishes have a central feed horn surrounded by an `inner-ring' of 6 feed horns. Parkes has an `outer-ring' of 6 additional feed horns. To compute the response of a feed-horn to a point-like radiator of spherical waves, we first compute the electric field on the dish surface. We employ Huygens' principle and treat each segment of the dish as a secondary spherical radiator. We then sum up the electric fields of the ensuing spherical waves at each point on the focal plane. We finally average the aggregate electric field on the focal plane, over the aperture of the horn. The final averaging step gives us the response of the fundamental TE$_{11}$ mode of the horn to unpolarized radiation. By varying the position of the radiator, we can evaluate the response of any feed-horn to near and far field events occurring at varying angular positions with respect to the telescope's boresight. We do not consider the response to polarized signals in this paper. \\

\begin{figure*}
\centering
\includegraphics[width=\linewidth]{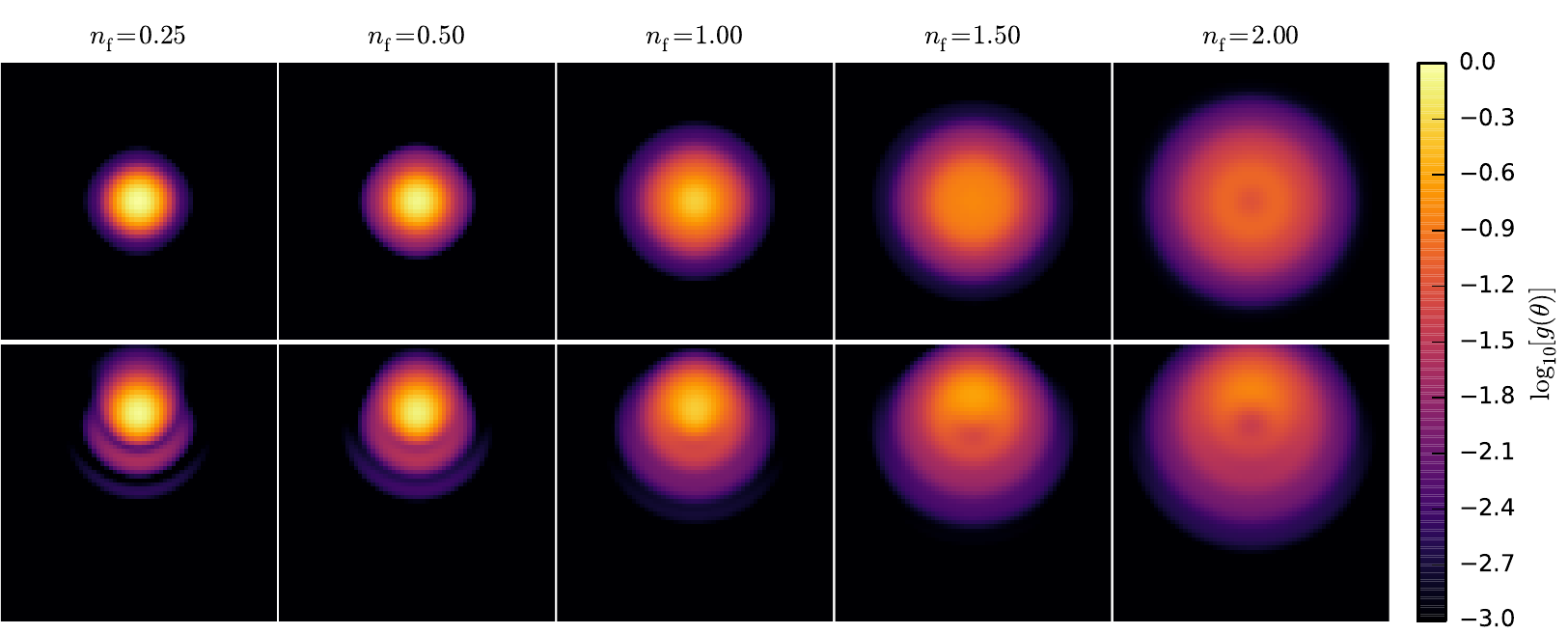}
\caption{Simulated beam patterns of a dish antenna at varying distances $D$ to a point-like radiator. Distances are specified in terms of their Fresnel number $n_{\rm f} = \frac{d^2}{4\lambda D}$. Top row: central feed; bottom row: inner ring feed.\label{fig:beams}}    
\end{figure*}

Fig.~\ref{fig:beams} shows a set of beams for the central feed (top row) and an inner-ring feed. The columns represent the near-field response at varying distances $D$ from the dish, indexed here in terms of the Fresnel number\footnote{Fresnel number is the phase difference, in units of $\pi$, of the incident field between the center of the aperture to its edge.} $n_{\rm f} = \frac{d^2}{4\lambda D}$, where $d$ is the dish diameter, and $\lambda$ is the wavelength. The $n_{\rm f}=0.25$ beams are representative of the far-field response to good accuracy, and for increasing $n_{\rm f}$ (decreasing $D$) the beams get progressively defocused. The far-field FWHM of the central beam in our simulation for the Parkes dish is about 14'.5 at 1.4~GHz which agrees with the quoted value of 14'4 to better than 5\%. The first coma lobes for the inner and outer ring feeds in our simulation are respectively at 18~dB and 13.85dB below the peak gain values. The departure of the coma lobe levels from quotes values of 17~dB and 14~dB is less than $20$\% and $5$\% respectively. \\

\subsection{Probability of multiple-beam events}
\label{sec:pmbe}
The probability of multiple-beam events depends on the number of neighboring beams at different points on the focal plane. Anticipating this dependence, we partition the sky into as many regions as the number of beams. Each region then has a `principal' beam which will register the highest flux-density among all beams were a burst to occur in that sky region. The burst may be additionally detected in one or more auxiliary beams. 
We consider integral source counts of the form in Equation~\ref{eqn:lnls}. We assume that the same detection threshold is applied to the data streams from all beams. We absorb any inter-beam variation in system temperature and aperture efficiency into the antenna-beam gain (see Appendix A1). The probability of detecting a burst coming from a solid-angle element $\delta^2\mv{l}$ at an angular offset $\mv{l}$ from a beam's boresight is then $\mathcal{P}(\mv{l}) \propto g^{\alpha}(\mv{l})\delta^2\mv{l}$, where $g(\mv{l})$ is the antenna beam gain towards direction $\mv{l}$.\\

To compute the probabilities of multiple-beam events, we use the following algorithm.
\begin{enumerate}
\item For each pixel $\mv{l}$ in the sky, sort the gains of the beams towards that pixel in decreasing order: $[g_1(\mv{l}),\, g_2(\mv{l}),\, ...]$ etc. Here, $i=1$ is the principal beam by construction.
\item The threshold fluence for detection in the $i^{\rm th}$ beam is proportional to $g^{-1}_i(\mv{l}) $. The probability of a $n$-beam detection at pixel $\mv{l}$ is thus 
\begin{eqnarray} 
\mathcal{P}(n,\mv{l}) \propto & g^{\alpha}_n(\mv{l})-g^{\alpha}_{n+1}(\mv{l});& n<n_{\rm beam}\nonumber \\
                                  & g^{\alpha}_{n};& n=n_{\rm beam}
\end{eqnarray}
where $n_{\rm beam}$ is the number of feed-horns.

\item Marginalize over $\mv{l}$ to get $\mathcal{P}(n) = \int {\rm d}^2\mv{l}\mathcal{P}(n,\mv{l})$. We set the normalization to get $\sum_{i=1}^{i=n_{\rm beam}}\int {\rm d}^2\mv{l}\mathcal{P}(i,\mv{l}) = 1$, such that all probabilities computed are conditional upon a burst being detected. 

\item The probabilities for cases where a particular beam is chosen {\it a priori} as the principal beam can be computed by only integrating over the sky pixels that belong to the beam's partition.
\end{enumerate}
In Table~1, we present the probabilities for Parkes multiple-beam detections for far-field events. These probabilities were computed using frequency-averaged simulated beam patterns over the range 1182--1525~MHz. Corresponding probabilities for near-field events are given in the Appendix (Fig.~\ref{fig:prob13}).

\begin{table}
\centering
\label{tab:mbp}
\caption{Multiple-beam detection probabilities for the Parkes multi-beam receiver when the source is in the far field. The 2- and 3-beam detection probabilities for the Euclidean value of $\alpha=-1.5$ is very low: $\lesssim$ 1 in 300, 150 and 230 events for central inner-ring and outer-ring feeds respectively.} 
\begin{tabular}{llll}
\hline
Feed,$\alpha$	& 1-beam	& 2-beam	& 3-beam\\ \hline \\
Central, $\alpha=0.5$   & 0.7839        & 0.0452        & 0.0476 \\
Central, $\alpha=1.0$   & 0.9608        & 0.0146        & 0.0116 \\
Central, $\alpha=1.5$   & 0.9935        & 0.0033        & 0.0019 \\ \hline
Inner, $\alpha=0.5$     & 0.7506        & 0.1074        & 0.0530 \\
Inner, $\alpha=1.0$     & 0.9542        & 0.0311        & 0.0099 \\
Inner, $\alpha=1.5$     & 0.9918       & 0.0065        & 0.0014 \\ \hline
Outer, $\alpha=0.5$     & 0.8409        & 0.0155        & 0.0155\\
Outer, $\alpha=1.0$     & 0.9775        & 0.0194        & 0.0014 \\
Outer, $\alpha=1.5$     & 0.9954        & 0.0044        & 0.0001\\ \hline\\
\end{tabular}
\end{table}
As seen from Table~1, the expected number of multiple-beam detections from a non-evolving population in Euclidean space ($\alpha=1.5$) is very low:  $\lesssim$ 1 in 300, 150 and 230 events for central inner-ring and outer-ring feeds as their principal beams respectively (95\% confidence). The Parkes beams are spaced further apart than their half-power widths which leads to such low probabilities for multiple-beam detections. These numbers are in stark contrast with 2 in 16 events seen at Parkes in two or more beams.   
\section{Parkes multi-beam constraints on $\bm{\alpha}$}
\label{sec:alpha_mbs}
We will now compute the value of $\alpha$ that  best satisfies the rate of multiple-beam detections among FRBs discovered at the Parkes telescope. If the probability of detecting a burst in $i$ beams is $\mathcal{P}(i)$, then the probability of detecting $k$ bursts out of $n$ in $i$ beams is then given by the binomial distribution
\begin{equation}
\mathcal{P}[k\textrm{ in }n;\, i\textrm{ beams}] = {n \choose k} \left[\mathcal{P}^k(i)\left(1-\mathcal{P}^{n-k}(i)\right)\right],
\end{equation} 
where 
\begin{equation}
{n\choose k} = \frac{n!}{k! (n-k)!}
\end{equation}
is the number of ways of picking $k$ unordered items from $n$ possibilities. 
In reality, the probability $\mathcal{P}(i)$ depends on the principal beam, since different feeds have different numbers and orientations of neighboring feeds. Accounting for this dependence is telescope specific. The computation for the FRBs detected at Parkes is given below.

All 15 published FRBs observed at Parkes are cataloged by \citet{frbcat}. Of these events, 14 were reported as single-beam detections\footnote{We encourage the discoverers of FRBs to always report on adjacent-beam constraints.}; 2, 8, and 4 detections had their principal beam corresponding to the central, inner, and outer ring feeds respectively. One \citep{lorimer2007} was detected in 4 beams, with an inner-ring providing the principal detection. An additional burst (Ravi et al., in prep) was detected in two beams, again with an inner-ring principal beam. We defer an analysis of the probability of a given feed to be the principal beam to future work. In this paper, we take the principal beams for each event as given. 

\begin{figure}
\centering
\includegraphics[width=\linewidth]{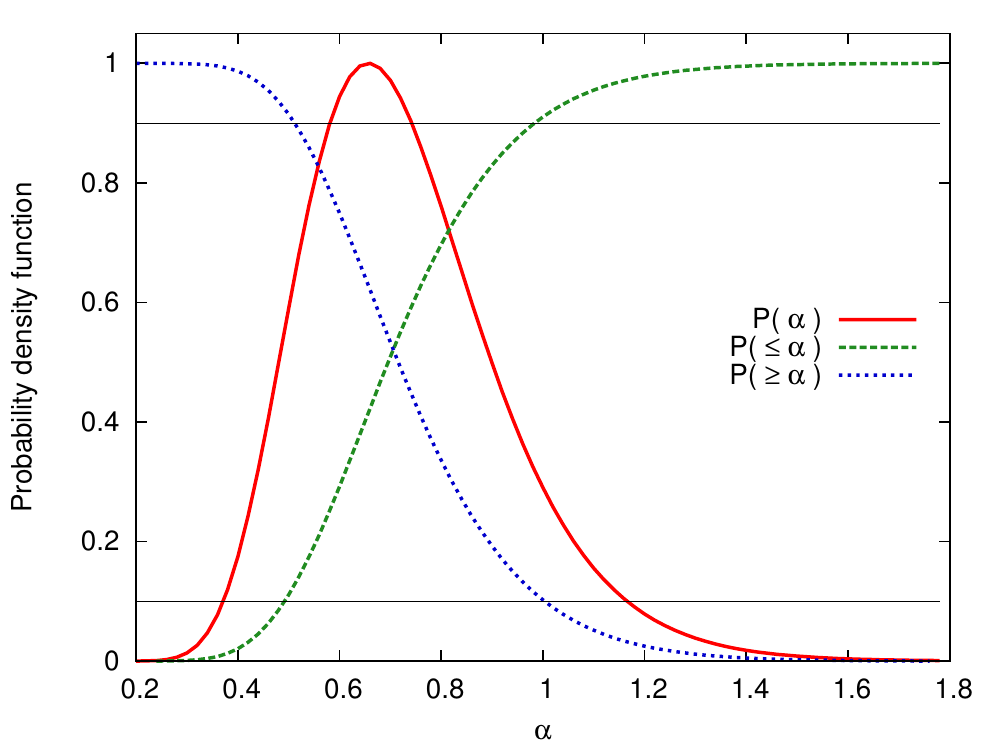}
\caption{Posterior probability distribution of $\alpha$ given the 16 Parkes detections (solid red curve), conditional upon FRBs being far-field events with integral source counts given by $N(>\fobs) \propto \fobs^{-\alpha}$. The green dashed curve is the corresponding cumulative distribution function, and the blue dotted line is its complement. The posterior bounds on $\alpha$ are $0.52<\alpha<1.0$ at 90\% confidence. \label{fig:alpha_modsel}} 
\end{figure}

The aggregate probability of achieving these 16 Parkes detections is:
\begin{eqnarray}
\mathcal{P}(\textrm{parkes 16} | \alpha) &=& {16 \choose 8}\mathcal{P}^8_{\rm i}(1) \times {8 \choose 4}\mathcal{P}^4_{\rm o}(1)\nonumber \\
&& \times {4 \choose 2}\mathcal{P}^2_{\rm c}(1) \times \mathcal{P}_{\rm i}(4)\nonumber \\
&& \times \mathcal{P}_{\rm i}(2)
\end{eqnarray}
where a dependence on $\alpha$ of all the probabilities on the right hand side is implicitly assumed for brevity, and the subscripts denote the position of the principal beam. We now assume that FRBs are all far-field events, i.e., they originate from $n_{\rm f}\lesssim 0.25$ or equivalently from distances $d\gtrsim 20$~km. The posterior probability distribution of the 16 Parkes detections for various values of $\alpha$ as
\begin{equation}
\mathcal{P}(\alpha | \textrm{Parkes 16}) = \frac{\mathcal{P}(\textrm{Parkes 16} | \alpha) \mathcal{P}(\alpha)}{\mathcal{P}(\textrm{Parkes 16})}
\end{equation} 
Being agnostic about the FRB progenitors, we choose a flat prior on $\alpha$ in the (unconstraining) range $0.2<\alpha\leq1.8$, and by restricting ourselves to only models with variations on $\alpha$, we can evaluate the evidence in the denominator as
\begin{equation}
\mathcal{P}(\textrm{Parkes 16}) = \sum_i \mathcal{P}(\textrm{Parkes 16} | \alpha_i)\mathcal{P}(\alpha_i).
\end{equation}
Fig.~\ref{fig:alpha_modsel} shows the posterior probability of $\alpha$ given the 16 Parkes detections. Very low and very high values of $\alpha$ are disfavored by the relative paucity and abundance respectively of multiple-beam detections. The 90\% confidence bound on $\alpha$ is given by  $0.52<\alpha <1.0$, which is significantly different from the value of $\alpha=1.5$ expected for a non-evolving population in Euclidean space. We reject $\alpha=1.5$ with $>99\%$ confidence.\\

We have evaluated the robustness of our results against survey incompleteness at faint fluence levels where the bursts are predominantly expected to be single-beam events. \citep{keane2015} studied survey incompleteness effects in simulations and found that up to 22\% of the bursts can be missed for a Euclidean distribution of FRB fluences ($\alpha=1.5$). This corresponds to 5 missed detections. We recomputed the bounds on $\alpha$ for a hypothetical scenario where 5 more FRBs are discovered at Parkes. We assume that all 5 are single-beam detections, 3 of which are detected in an inner ring beam and the remaining 2 in an outer ring beam. The constraint on $\alpha$ for this hypothetical scenario is $0.58<\alpha<1.06$ at 90\% confidence, which clearly demonstrates the robustness and unbiased nature of our bounds on $\alpha$ against survey incompleteness. In addition, our results are robust to variations in intrinsic burst properties since they do not affect the \emph{fraction} of multiple-beam detections as considered here. \\

We finally note that our bounds are, as expected, highly sensitive to the number of multiple-beam detections. For instance, dropping the 4-beam event FRB\,010724 \citep{lorimer2007} from our calculations revises the constraint to $0.88<\alpha<1.52$. This is still marginally inconsistent with  a Euclidean population.
\section{Multi-telescope detection statistics}
\label{sec:alpha_mt}
Several telescopes have been searching for FRBs, each with their own sensitivities and survey volumes. The number of expected FRB detections from a given telescope depends on the nature of the source count distribution (log$N$-log$F$ curve) for FRBs. For instance, larger dishes have more collecting area but narrower fields of view. They are thus best suited to detecting faint events, and are favored by a steep log$N$-log$F$ slope that implies a large number of faint objects. Shallower log$N$-log$F$ distributions, on the other hand, yield large numbers of bright events which makes the telescope sensitivity less relevant than the field of view in achieving greater detection rates. Hence, detections, or the lack thereof, from telescopes with varying sensitivities and fields of view carry important information about that the nature of FRB source counts. Motivated by this, we will now derive the posterior probability distribution for the source-count parameter $\alpha$. 

\begin{table*}
\centering
\caption{List of telescope parameters for FRB searches at facilities other than Parkes. $d$: dish diameter, $T_{\rm sys}$: system temperature, $\eta_{\rm eff}$: aperture efficiency, $\Delta\nu$: observation bandwidth, $\tau_{\rm ds}$: dispersion smearing timescale, $N_{\rm ant}$: number of antennas, $N_{\rm day}$: number of days of exposure, $\zeta$: detection threshold in units of thermal noise.}
\begin{tabular}{llllllllll}
\\ \hline
Name	& $d$~[m]	& $T_{\rm sys}$~[K]	& $\eta_{\rm eff}$	& $\Delta\nu$~[MHz]	& $\tau_{\rm ds} [ms]$ & $N_{\rm ant}$	& $N_{\rm day}$ &  $\zeta$ & Reference\\ \hline \\
ATA 	& 6.0		& 92	& 0.6	& 210	& 3.9	& 1.0	& 16.475 & 5.5	& \citet{siemion2012}\\
ARC		& 220.0		& 50	& 0.6	& 322.6	& 0.8	& 1.0	& 82.6	& 7.0	& \citet{spitler2014}\\
VLA		& 25.0		& 50	& 0.6	& 256	& 2.4	& 27	& 6.917	& 7.5	& \citet{law2015}\\
AS1		& 0.1		& 850	& 1.0	& 590	& 2.7	& 1		& 285 	& 6.0	& \citet{saint2014}\\
AS2		& 1.2		& 850	& 1.0	& 590	& 2.7	& 1		& 591.7	& 6.0	& \citet{saint2014} \\\hline
\end{tabular}
\label{tab:survey_params}
\end{table*}

Although the Parkes constraint on $\alpha$ was independent of the normalization of the source counts, this is not the case here. 
We assume the following integral source-counts in this section \citep{law2015}:
\begin{equation}
\label{eqn:eventrate}
\mathcal{N}(>\fobs) = \frac{1.2\times 10^4}{4\pi}\left(\frac{\fobs}{1.8\textrm{~Jy~ms}}\right)^{-\alpha}\textrm{sr$^{-1}$day$^{-1}$},
\end{equation}
and consider the sensitivity of our results to variations in the source-count normalization in Section~4.2. Table~2 summarizes the various telescope parameters for the different published results we consider here; for details, see Appendix~A2. To design a common algorithm to compute the necessary statistics, we have absorbed telescope efficiency parameters into the system temperature, $T_{\rm sys}$, so that the final thermal noise per time-integration in our formalism matches the values quoted.\\

\subsection{Detection rates}
We assume that FRBs have a mean duration of $\tau_{\rm FRB}=3$~ms, a mean DM of 780~pc cm$^{-3}$ \citep{law2015}, and 
no correlation between the two quantities. If $\Delta\nu$ is the bandwidth, $d$ is the aperture diameter, $\eta$ is the aperture efficiency, and 
$\tau_{\rm int}$ is the spectrometer integration time, then the thermal noise flux-density in a single time-integration for an incoherent summation of 
signals from $N_{\rm ant}$ antennas is 
\begin{equation}
\label{eqn:th_noise}
S_{\rm th} = \frac{8k_{\rm B}T_{\rm sys}}{\eta\pi d^2}\frac{1}{\sqrt{2N_{\rm ant}\Delta\nu\tau_{\rm int}}},
\end{equation} 
where $k_{\rm B}$ is Boltzmann's constant. The threshold for detection depends on the amount of dispersion smearing, the temporal width of the burst with respect to the integration time, and the threshold used for detection (number of $\sigma$ above thermal noise) $\zeta$. The fluence threshold for detection can then be written as
\begin{equation}
\mathcal{F}_{\rm det}(\theta) = S_{\rm th}\tau_{\rm FRB} \,\frac{\zeta}{r_1r_2g(\theta)}
\end{equation}
where $g(\theta)$ is the power gain of the telescope aperture for an angular offset $\theta$ from boresight. The factors $r_1$, and $r_2$ approximately account for dilution of FRB fluence due to time integration, and SNR boost due to the number of independent epochs combined during a detection. They are respectively given by
\begin{equation}
r_1 = \left[\frac{\mathcal{M}(\tau_{\rm FRB},\tau_{\rm ds})}{\tau_{\rm int}} \right]_0^1
\end{equation}
\begin{equation}
r_2 = \left[\sqrt{\frac{\mathcal{M}(\tau_{\rm FRB},\tau_{\rm ds})}{\tau_{\rm int}}} \right]_1^\infty
\end{equation}
where the subscript and superscript in $\left[. \right]_a^b$ represent the lower and upper bounds for the values within the square brackets, and the function $\mathcal{M}(.)$ yields the largest of its arguments. If a survey observes for $N_{\rm day}$ days, then the expected number of detections is given by.
\begin{equation}
\label{eqn:ndet}
N_{\rm det} = N_{\rm day}\int {\rm d}\theta 2\pi\sin\theta\, \mathcal{N}\left( > \mathcal{F}_{\rm det}(\theta)\right) 
\end{equation}    
where $2\pi\sin\theta{\rm d}\theta$ is the differential solid angle. Finally, while computing the multi-telescope detection statistics, we will assume that $g(\theta)$ is given by the Airy function:
\begin{equation}
\label{eqn:airy}
g(\theta) = \left(2\,\frac{J_1(\pi d/\lambda \sin\theta)}{\pi d/\lambda\sin\theta} \right)^2.
\end{equation}
As all surveys that we consider operate in approximately the same frequency bands, the effects of frequency-dependent scatter-broadening are 
absorbed into the assumed burst width.  

\subsubsection{Constraints on $\alpha$}
We now use Equation~\ref{eqn:eventrate} along with the survey parameters mentioned in Table~2 to constrain $\alpha$. While doing so, we are invariably extrapolating the source counts computed from one fluence regime to another since different telescopes have different detection thresholds. We must thus carefully consider possible turn-overs or cut-offs in the source population towards large fluences.\\

As will be shown in Section~\ref{subsec:optimum_dish}, for $\alpha<1$, a survey with a smaller dish (larger FOV) will detect more events as compared to one with a larger dish. Since the Parkes multiple-beam detection rates imply $\alpha\lesssim 1$, it is important to consider the (non-)detections from the ASSERT survey \citep{saint2014}, which among published rate-limits at L-band has both the largest exposure time and the smallest dish. Since ASSERT found no FRBs, we make the reasonable assumption that there is a maximum FRB fluence cut-off at about $50$\,kJy\,ms--- consistent with the sensitivity of the ASSERT program. In addition, the inferred fluence of the brightest event observed thus far sets a lower limit on the maximum cut-off fluence. Based on modeling of the `Lorimer burst' event at Parkes \citep{lorimer2007}, the intrinsic fluence of the brightest Parkes burst is expected to be, at most, about 500~Jy~ms (Ravi et al., in prep.). We thus marginalize all probabilities derived in this section over the cut-off fluence while assuming a uniform prior between 0.5 and 50~kJy~ms.\\

\begin{figure}
\centering	
\includegraphics[width=\linewidth]{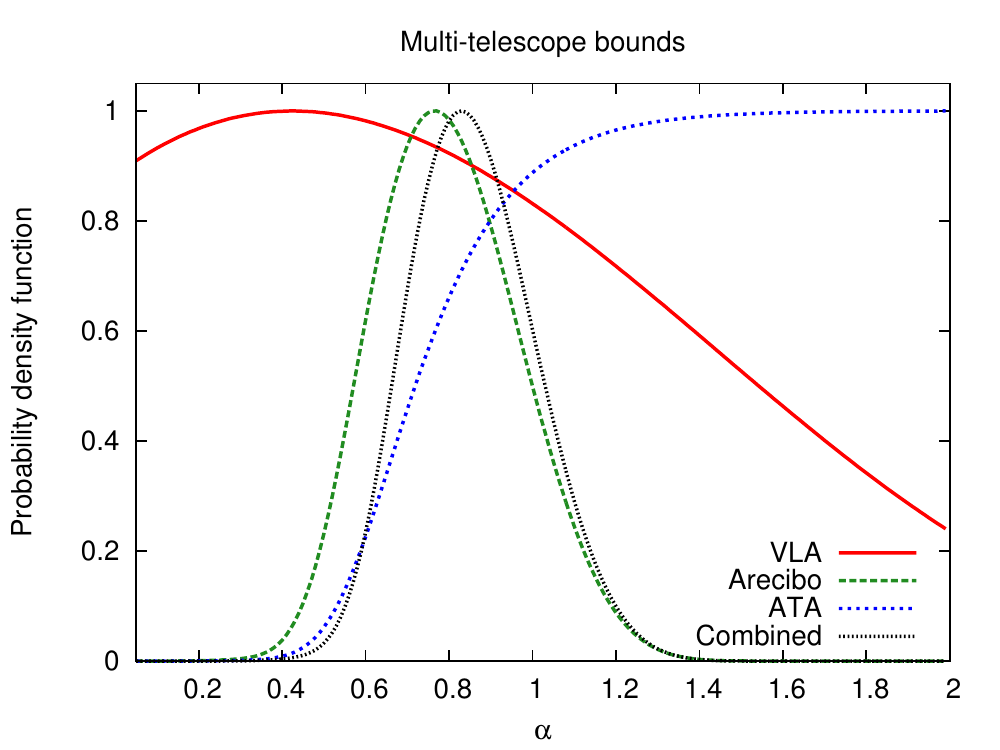}
\caption{Bounds on log$N$-log$F$ parameter $\alpha$ imposed by detections and non-detections at other facilities with different aperture diameters and sensitivities: $0.67<\alpha<1.07$ (90\% confidence). The probabilities have been marginalized with respect to the maximum cutoff fluence $\mathcal{F}_{\rm cut}$.\label{fig:alpha_bounds_multitel}}
\end{figure}

Fig.~\ref{fig:alpha_bounds_multitel} shows the probability density function of $\alpha$ evaluated using Equation~\ref{eqn:ndet} for the various surveys whose parameters are given in Table~2. In doing so we have assumed Poisson statistics for the arrival of FRBs, i.e if the expected numbers of events for a survey is $N_{\rm det}$, then the probability of discovering $N_{\rm event}$ events in a survey is
\begin{equation}
\mathcal{P}(N_{\rm event},N_{\rm det})  = \frac{N_{\rm det}^{N_{\rm event}} {\rm e}^{-N_{\rm det}}}{N_{\rm event}!}.
\end{equation} 
As seen in the Figure, the strongest constraints on $\alpha$ come from the Arecibo telescope, owing to its excellent sensitivity afforded by the large collecting area. For very flat log$N$-log$F$ distributions ($\alpha\lesssim 0.5$), we expect a large number of bright events which will be detected even in the sidelobes of the Arecibo's beam pattern. This partially offsets the small FOV of the Arecibo dish. Steeper log$N$-log$F$ distributions ($\alpha\gtrsim 1.1$) simply yield a large number of faint events which will cross Arecibo's detection threshold around boresight. The VLA also has a large collecting area, but since its FRB search is only restricted to the FWHM of the primary beam, the VLA non-detections cannot rule out low values of $\alpha$. The Allen Telescope Array (ATA) on the other hand is only sensitive to relatively bright events that occur close to its boresight, and the non-detections from the ATA can only rule out very flat ($\alpha\lesssim 0.6$) log$N$-log$F$ distributions. \\

Since the surveys are independent trials, we multiply their respective probabilities for (non-)detections to get the aggregate probability. The posterior probability of $\alpha$ is then computed by assuming an evidence that normalizes the integral of the the aggregate probability to unity. The multi-telescope constraints on $\alpha$ thus obtained are $0.67<\alpha<1.07$ at 90\% confidence, which is in excellent agreement with the Parkes multiple-beam detection constraint. Multiplying the Parkes multiple-beam and multi-telescope probability density functions, we obtained our final constraint of $0.66<\alpha<0.96$ at 90\% confidence.\\

\subsection{Some caveats}
We caution the reader that the multi-telescope constraints may suffer from certain systematic errors. As pointed out by \citet{law2015}, the source count assumed here (Equation~\ref{eqn:eventrate}) has been estimated based on average burst properties such as DMs, intrinsic widths, scattering timescales etc. The distribution functions for these properties are not well known. In addition, the constraints on $\alpha$ are somewhat degenerate with the over-all normalization of the all-sky FRB rate\citep{oppermann2016}. To gauge the sensitivity of our constraints to such effects in a simplified manner, we have recomputed the confidence interval for $\alpha$ for a selection of cases. We have assumed the fiducial values for DM, $\tau_{\rm FRB}$ and the normalization of FRB source counts of 780~pc~cm$^{-3}$, 3~ms, and $1.2\times 10^4$ events above a fluence of 1.8~Jy~ms per day respectively. In each case, we vary one of these three parameters by 100\% while fixing the other to their fiducial values. 
\begin{itemize}
\item Consider the number of events per day above a fluence of 1.8~Jy~ms to be $0.6\times10^4$ or $2.4\times 10^4$ (see Equation~\ref{eqn:eventrate}). The respective constraints on $\alpha$ are  $0.63<\alpha<1.16$ and $0.71<\alpha<0.99$. 
\item Consider the mean FRB width to be $\tau_{\rm FRB}=1.5$~ms or $\tau_{\rm FRB}=6$~ms. The respective constraints on $\alpha$ are $0.66<\alpha<0.99$ and $0.68<\alpha<1.15$.
\item Consider the mean DM to be 375~pc cm$^{-3}$ or 1600~pc cm$^{-3}$. The respective constraints on $\alpha$ are $0.68<\alpha<1.07$ and $0.66<\alpha<1.06$. 
\end{itemize}
Hence, our constraints are robust to even $100$\% changes in the assumed FRB-rate normalization, mean FRB width and DM. We do however note that a drastic reduction in the all sky FRB rate by a factor of $\sim 10$ yields values of $\alpha$ that are roughly consistent with a Euclidean population.\\ 

In addition to burst properties, there may be systematic effects due to practical choices in experimental design and detection algorithms. The Arecibo search for FRBs for instance, was limited to low Galactic latitudes where observed FRB fluences may be significantly lower due to scintillation-induced biases \citep{macquart2015}.\footnote{Scintillation is not expected to change the log$N$-log$F$ slope for a fluence-range away from any cut-offs \citep{macquart2015}.} In addition, since the Arecibo FRB search was limited to DM less than 2000~pc~cm$^{-3}$ \citep{spitler2014}, weaker events that preferentially originate from larger distances may have been overlooked. Finally, as single-burst detection techniques are evolving, different surveys (even on the same telescope) may be employing algorithms with different missed-detection and false-positive rates which makes it difficult to bring their (non-)detections into a common probabilistic framework. Nevertheless, the robustness of our constraints on $\alpha$ against large variations in the event-rate normalization and mean FRB characteristics lends credibility to our results despite these misgivings. 
\section{Discussion}
\label{sec:disc}
The Parkes multiple-beam detection rates and the non-detections at other facilities strongly favor an FRB distribution that has a remarkably flat log$N$-log$F$ distribution: $0.66<\alpha<0.96$ (90\% confidence) as compared to that expected in a Euclidean Universe $(\alpha=1.5$) with a non-evolving source population. This has implications both for design of future surveys and for theories regarding the progenitor population. We discuss these aspects below.
\subsection{Implications for survey design}
\label{subsec:optimum_dish}
\begin{figure*}
\includegraphics[width=\linewidth]{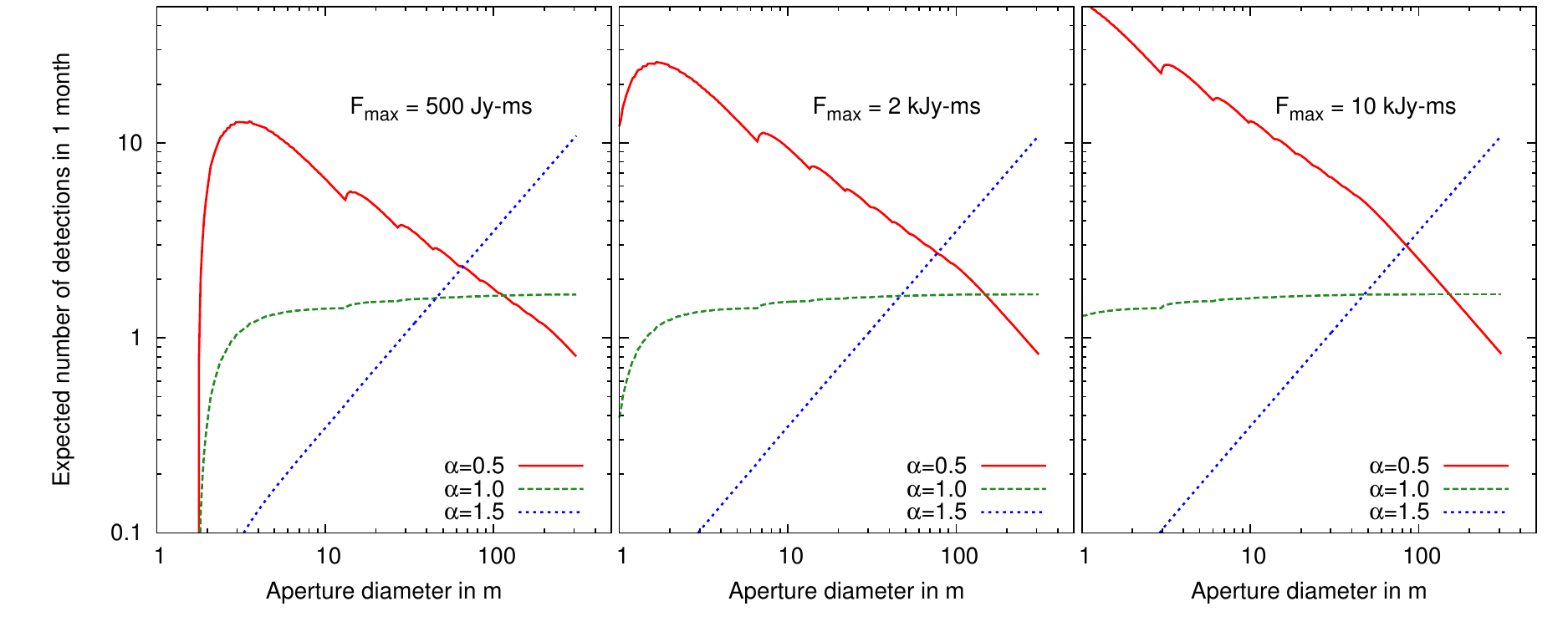}
\caption{Expected number of detections in a month from a hypothetical 10-element FRB array as a function of antenna dish diameter. We have assumed $T_{\rm sys}=60$~K, $\eta=60$\%, $\Delta\nu 300$~MHz, $\nu=1.5$~GHz, an event rate given by Equation~\ref{eqn:eventrate}, and that the detection is made on the incoherent sum of signals from the 10 dishes. The panels correspond to different maximum fluence-cutoffs that current observations allow. The red-solid, green-dashed, and blue-dotted lines correspond respectively to a log$N$-log$F$ parameter, $\alpha$, of 0.5 1.0, and 1.5 respectively. Based on our bounds of $0.66<\alpha<0.96$ we conclude that hypothetical array will detect (and localize) at least 1 FRB per month. \label{fig:eventrate}} 
\end{figure*}
We will now compute the observed number of FRB-like events for a hypothetical array as a function of dish size (single pixel receiver). Our aim is to determine the `optimum' dish-diameter to maximize the number of detections. Fig.~\ref{fig:eventrate} shows the number of detections per month computed using Equations~\ref{eqn:th_noise} to~\ref{eqn:airy} for a (hypothetical) array of $10$ dishes whose outputs are incoherently combined to detect FRBs. We assume the following parameters: $T_{\rm sys}=60$~K, $\eta=0.6$, $\nu=1.5$~GHz, $\Delta\nu=500$~MHz, $\tau_{\rm ds}=1$~ms, $\tau_{\rm FRB} = 3$~ms, $\zeta=8$. The different curves are for different values of the log$N$-log$F$ slope parameter $\alpha$. We assume that the source-counts cut-off at fluence $\mathcal{F}_{\rm max}$, and that the source-count normalization is given by Equation~\ref{eqn:eventrate}. \\

Clearly, $\alpha=1$ is the dividing line between the FOV and sensitivity domains: $\alpha>1$ yields a paucity of bright events and larger, more sensitive telescopes win. Brighter events are relatively plentiful for $\alpha<1$ which favors smaller dishes with larger fields of view. If we conservatively assume that $\alpha=1.0$ and that the maximum cut-off fluence is 500~Jy~ms (see Fig.~\ref{fig:eventrate}), then the optimal dish diameter is $d\sim 6$~m--- a value at which the $\alpha=1$ line begins to saturate. Smaller dishes may be insensitive to a large number of events, and significantly larger dishes will have excluded large numbers of events due to their narrow fields of view. For the most likely range of $0.66<\alpha<0.96$, we find that dish diameters of between 
1\,m and 6\,m are preferred, and that the detection rate could be well over 10 events per month. Hence, we conclude that given the constraints on $\alpha$ presented here, a modest array ($N_{\rm ant}\sim 10$) of small dishes of about $d\sim 6$~m will detect at least $\gtrsim 1$ FRB per month. Future FRB surveys may take advantage of this fact and design for a system that detects events using the incoherent sum of the dish spectra, and dump raw voltages (written in real time to a circular buffer) for interferometric localization post-detection. The ATA with its 6-meter dishes may benefit greatly from the implementation of such a detection and localization strategy.\\

\begin{figure}
\centering
\includegraphics[width=\linewidth]{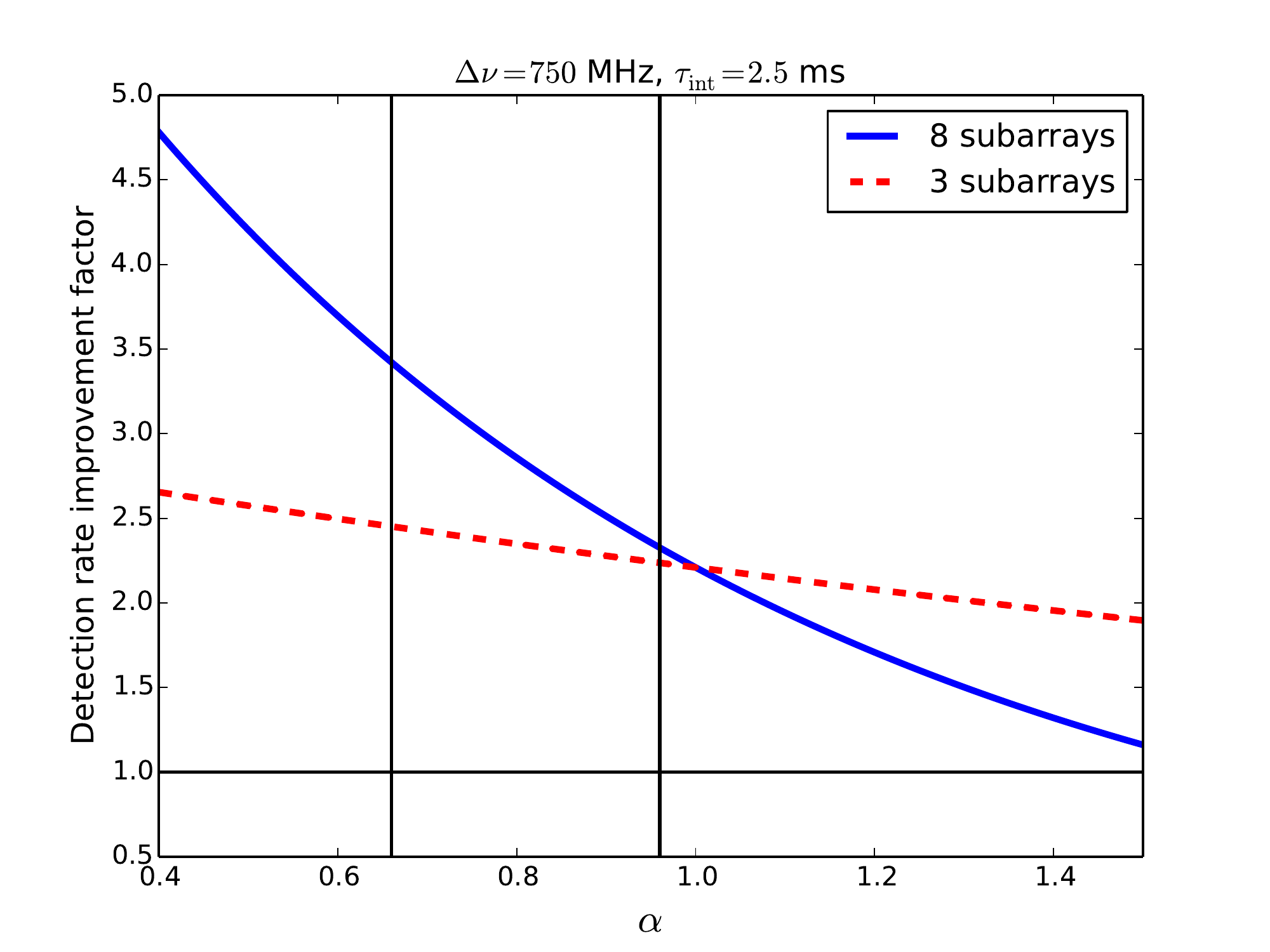}
\caption{Improvement in the detection rate expected with the subarray-mode at the VLA for 3 subarrays (broken red line) and 8 subarrays (solid blue line) as a function of the log$N$-log$F$ parameter $\alpha$. The improvement is computed over the observation mode used by \citet{law2015} ($\Delta\nu=256$~MHz, $\tau_{\rm int}=5$~ms). The thin vertical lines represent the 90\% confidence bounds of $0.66<\alpha<0.96$ from this work. We have assumed that the reduction in the total number of baselines due to subarraying will allow for a larger bandwidth $\Delta\nu=750$~MHz, and a smaller integration interval of $\tau_{\rm int}=2.5$~ms.\label{fig:vla_subarrays}}     
\end{figure}

We next consider the FRB discovery and localization program at the VLA \citep{law2015} as a `case in point' for how our bounds on $\alpha$ can have a significant impact on survey design. Consider partitioning the 27 antennas of the VLA into `subarrays'--- groups of antennas that operate as independent interferometers, each with a unique pointing center.\footnote{The Fly's Eye search at the ATA is a special case of the subarrays concept where each subarray has a single primary antenna element.} Subarraying is essentially a FoV--sensitivity trade-off, and since we find $\alpha<1$ with 95\% confidence, the expected number of detections improves with increasing number of subarrays. In addition, since the data rate of an interferometer with $N$ elements, scales as $N^2$, the data rate for $N_{\rm sub}$ subarrays scales as $N^{-1}_{\rm sub}$. This reduction in data rate opens up the possibility to employ larger bandwidths and shorter correlator integration-times further improving the sensitivity to detection. In Fig.~\ref{fig:vla_subarrays}, we compute the improvement in detection rates that we expect by using subarrays at the VLA. We have assumed that the reduced number of baselines will allow for an increased bandwidth of $\Delta\nu=750$~MHz and a reduced integration time $\tau_{\rm int}=2.5$~ms, as compared to the values of $\Delta\nu=256$~MHz, and $\tau_{\rm int}=5$~ms used by \citet[][(no sub-arraying)]{law2015}. We find that given our constraints on $\alpha$, detection rates with the VLA can be increased three-fold by using 8 subarrays. 
\subsection{Implications for FRB distances}

\subsubsection{Euclidian-space calculation}

First, our bounds on $\alpha$ strongly disfavor models where FRBs are standard candles, since in that case, the log$N$-log$F$ function will have a slope of $\alpha=1.5$ barring carefully contrived source population evolution with distance (see Appendix~A3). 
Next, we consider cases where FRBs have an intrinsic burst energy distribution that is a power-law with index $\beta$, i.e $\mathcal{N}(>\fint) \propto \fint^{-\beta}$ where $\fint$ is the intrinsic burst energy.\footnote{We assume that any relativistic beaming effects are absorbed into $\fint$.} Note that if $\mathcal{N}(>\fint)$ does not evolve with distance, then the observed fluence distribution follows the   Euclidean values of $\alpha=1.5$ for any intrinsic energy distribution function $\mathcal{N}(>\fint)$. Considering $\beta$ and the distance evolution law as unrestricted, we can obtain a large range of values for $\alpha$, which we will not consider here, since we do not have good physical motivations to assume values for either of the factors. Instead, we will consider two limiting cases where there is a minimum and maximum cutoff distance ($\rmin$ and $\rmax$) to FRBs respectively. We will further assume that the observed population is affected by such cut-offs, failing which, the observed fluence distribution will revert to a Euclidean value. That is,  $\fint^{\rm min}>4\pi\rmin^2\fobs$ and/or $\fint^{\rm max}<4\pi\rmax^2\fobs$, where $\fint^{\rm min}$ and $\fint^{\rm max}$ are the minimum and maximum intrinsic energies of the FRB population. In such cases, under reasonable assumptions, we can show that $\alpha=\beta$ (proof in Appendix~A3).\\

In addition, the number of sources detected in a survey that are within a distance $R$ evolves as $R^{3-2\beta} = R^{3-2\alpha}$, which for our bounds of $0.66<\alpha<0.96$ yields an event rate that scales as $\mathcal{N}(<R)\propto R^{+1.7}$ to $R^{1.1}$. The corresponding differential source counts 
are given by
\begin{equation}
\label{eqn:rlaw}
\frac{{\rm d}\mathcal{N}(R)}{{\rm d}R} \propto R^{0.1} {\rm ~to~} R^{0.7} {\rm ~at~ 90\%~ confidence.}
\end{equation} 
This implies that the FRBs are preferentially detected from larger distances. This is in stark contrast with a non-evolving population in Euclidean space for which the differential source-counts scale as  $R^{-1}$. Note that this does not prove that FRBs originate at cosmological distances, because the $\mathcal{N}(<R)$ curve will eventually saturate at some $R_{\rm s}$ at which $\fint^{\rm max}\approx 4\pi\fobs R_{\rm s}^2$, and $R_{\rm s}$ cannot be uniquely determined from the observed log$N$-log$F$ curve alone. \\ 

\subsubsection{Cosmological effects}

\begin{figure}
\centering
\includegraphics[width=\linewidth]{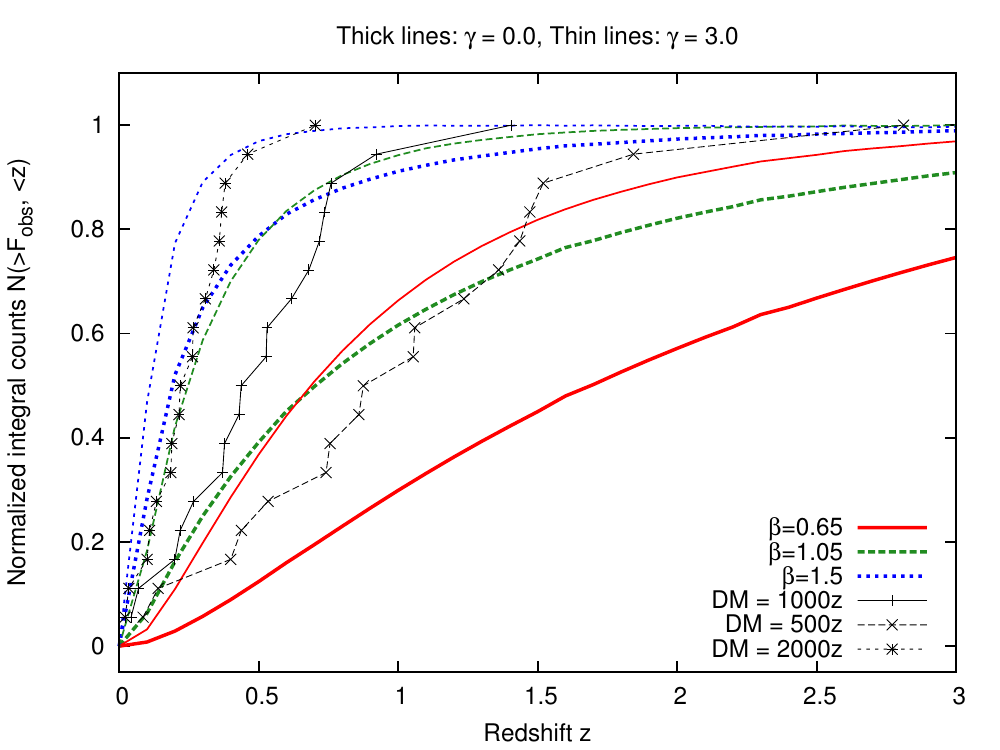}
\caption{Integral source counts $\mathcal{N}(<z)$ for a cosmological population of FRBs. The three curves are for $\beta=0.65$ and $\beta=1.05$ and $\beta=1.5$. The markers show the empirical cumulative FRB distribution where DM has been converted to redshift. 
Color correction has been applied with an FRB spectral index of $\gamma=1.0$. \label{fig:logN-logz}}
\end{figure}

Motivated by the above distance bias, we have recomputed the expected FRB fluence and distance distributions while taking cosmological effects into account (non-Euclidean geometry). For such a population we still find $\alpha=\beta$ but the distance-distribution is markedly different from the Euclidean-geometry case because of cosmological effects (Appendix~A4). Fig.~\ref{fig:logN-logz} shows the source counts for a cosmological population $\mathcal{N}(<z)$ for $\beta=0.65$, $\beta=1.05$ and $\beta=1.5$. The two sets of curves (thick and thin) are for intrinsic spectral indices of $\gamma=0.0$ and $\gamma=3.0$, where the observed and intrinsic fluence for a burst at redshift $z$ are related as $\fobs = \fint (1+z)^{-\gamma}$. The cumulative counts saturate at $z\gtrsim 1$ mainly due to a dramatic reduction in the rate at which the comoving volume element increases with redshift. This saturation is an important aspect of progenitor theories that place FRBs at cosmological distances, since it explains why a population of FRBs must come from a bounded volume despite the distance bias that is implied by our bounds on $\alpha$.\\ 

The black lines with markers in Fig.~\ref{fig:logN-logz} show the empirical cumulative distributions for the 17 published FRBs assuming different 
DM to redshift conversion factors. We have chosen the conversion factors to approximately span the uncertainty range that may be expected given simulations of the IGM baryon density structures \citep{dolag2015}. We only consider the excess DM over the expected Milky Way contribution in each case, and assume a host galaxy DM contribution of 50 pc~cm$^{-3}$. As is evident from the Figure, the uncertainties in the DM to redshift conversion and in the intrinsic burst spectral indices preclude us from evaluating which of the theoretical curves are best favored by the data. FRB localization and 
spectroscopic followup are required to definitively establish whether the FRB population adheres to the redshift-scaling implied by our constraints on 
$\alpha$.\\

As seen in Fig.~\ref{fig:logN-logz}, there is a clear paucity of events with ${\rm DM}\gtrsim1000$~pc~cm$^{-3}$, i.e the cumulative distribution saturates at DM~$\sim 1000$~pc~cm$^{-3}$. Unlike theories that place FRB progenitors at cosmological distances, progenitor theories that apportion the bulk of the extragalactic dispersion to the circum-burst media do not have a natural explanation for this apparent deficit of FRBs at ${\rm DM}\gtrsim$~1000. They are thus disfavored by our analysis. However, we are unable to make definitive statements on this point since high-DM events result in large burst durations prior to de-dispersion, and a sizable fraction may therefore be undetectable due to the greater chance of co-incident human-generated interference. We defer a detailed analysis of survey biases at high DMs to a future paper.\\

Finally, we caution the reader that in placing constraints on $\alpha$, we have implicitly made two assumptions: (i) FRB progenitors belong to a single family of objects, i.e., there is only one progenitor population, and (ii) the FRB fluence distribution is a power law with some index $\beta$, and some maximum cut-off fluence. It is often the case in astronomy that the diversity of objects whose emission adheres to some parameter space is not immediately apparent.\footnote{The discovery of sub-populations of gamma-ray transients is an example.} It is therefore entirely plausible, for example, that FRBs consist of two independent populations, with one being significantly brighter than the other. In this case, we may misconstrue events drawn from the aggregate as a common population with a flatter-than-usual log$N$-log$F$ law that shows an unusual propensity from brighter events. 

\subsubsection{Comparison with previous work}

Other authors \citep{manisha,li2016} have inferred log$N$-log$F$ slopes of $\alpha\approx 1$. \citet{li2016} discounted the effects of the primary beam by assuming that all FRBs are detected close to the antenna boresight, which may systematically bias their results towards larger values of $\alpha$. \citet{manisha} included the effects of the primary beam in their simulations and showed that a relatively large range of values of $\alpha$ ($0.9\pm0.3$) were consistent with the data. We note that, unlike previous studies, our Parkes multiple-beam constraints are largely insensitive and unbiased with respect to variations in burst properties due to scattering and dispersion.\\

Our results are at great odds with those of \citet{macquart2015} who infer that $\alpha>2.5$. \citet{macquart2015} invoked Galactic scattering to account for the apparent paucity of FRBs at low Galactic latitudes as reported by \citet{FRB010125}. \citet{FRB010125} found that even after accounting for Galactic scattering and dispersion, there is a deficit in low Galactic latitude detection rates, as compared to \citet{thornton2013}, at the 2.9$\sigma$ level. Its important to note that \citet{FRB010125} also found a discrepancy with an isotropic model at the 3.6$\sigma$ level. Since both models are rejected at high significance, a conservative interpretation of these results is that there is something other than FRB-latitude dependence that yields lower detection rates than those based on \citet{thornton2013}. In addition, $\alpha>2.5$ is strongly disfavored by the low detection rates with a large dishes such as Arecibo as shown in Fig. \ref{fig:alpha_bounds_multitel}. It is possible, however, that a majority of the faint events expected in the case of $\alpha>2.5$ are at DM$\gtrsim 2000$, which \citet{spitler2014} did not consider.

\section{Conclusions}

We have empirically constrained the fluence distribution (`log$N$-log$F$') of FRBs using two complementary approaches. Both (a) the probability of multiple-beam events registered by an array of receivers in the focal plane of a dish, and (b) the expected number of detections from dishes of different diameters, 
are dependent on the log$N$-log$F$ slope $\alpha$ (Equation~\ref{eqn:lnls}). We have combined these constraints on $\alpha$ to reach the following conclusions. All probabilities quoted below are computed with the prior assumption that the fluence distribution of FRBs is a simple power law with some maximum cut-off fluence. In addition, we have assumed that the detected population of FRBs has not been severely biased due to effects such as radio frequency interference, human errors (in evaluating candidate events) etc. 

\begin{itemize}

\item The incidence of multiple-beam events in the Parkes FRB sample constrains the log$N$-log$F$ slope to be $0.52<\alpha<1.0$ (90\% confidence).

\item The non-detections in FRB searches at the VLA, ATA, and ASSERT, together with the Arecibo FRB detection, yield $0.67<\alpha<1.07$ (90\% confidence). Taking this and the Parkes multiple-beam detection constraints together, we get $0.66<\alpha<0.96$ (90\% confidence).

\item The inconsistency of $\alpha$ with a value of 1.5 ($>99\%$ confidence) implies that (i) FRBs are not standard candles, and (ii) either the FRB luminosity distribution evolves strongly with distance, and/or FRBs progenitors are at cosmological distances. The former is disfavored by the relative paucity of FRBs at DM$\gtrsim 1000$~pc~cm$^{-3}$, although more work is needed to properly account for survey biases at high DMs, due to human-generated interference for instance. If FRBs are extragalactic, this result is inconsistent with a predominantly local-Universe population. 

\item If the intrinsic FRB cumulative energy distribution can be modeled as a power law with index $\beta$ ($\mathcal{N}(>\fint)\propto \fint^{-\beta}$), then  $\beta=\alpha$ which gives $0.63<\beta<0.95$ at 90\% confidence.

\item The expected number of detections of FRBs occurring at a distance $R$ scales as  ${\rm d}\mathcal{N}(R)/{\rm d}R\propto R^{0.1}$ to $R^{0.7}$ for our constrained range of $\alpha$ values. Hence distant events are either just as likely  or more likely to be detected by FRB surveys than nearby events. Note however that this does not prove that FRBs are at cosmological distances since the data are also consistent with $\mathcal{N}(>R)$ saturating at a non-cosmological  distance.

\item The relatively shallow slope of the FRB log$N$-log$F$ curve implies that a modest telescope array with $N_{\rm dish}\sim 10$ and $d\lesssim 6$~m aperture is sufficient to detect and localize a large population ($\gtrsim 1$~month$^{-1}$) of FRBs.  

\end{itemize}

\section*{Acknowledgments}
We thank Prof. Shrinivas Kulkarni for insightful discussions. We thank the CSIRO Australia Telescope National Facility for providing the engineering drawings of the 13-horn feed assembly on the Parkes telescope. HKV thanks Dr. Sarah Burke-Spolaor, Dr. Paul Demorest, and Dr. Casey Law for discussions regarding the concept of splitting the VLA into sub-arrays. 
\bibliography{mombib}

\begin{thebibliography}{}

\bibitem[\protect\citeauthoryear{{Burke-Spolaor} \&
  {Bannister}}{{Burke-Spolaor} \& {Bannister}}{2014}]{FRB010125}
{Burke-Spolaor}, S.,  \& {Bannister}, K.~W. 2014, \apj, 792, 19

\bibitem[\protect\citeauthoryear{{Caleb} et~al.}{{Caleb}
  et~al.}{2015}]{manisha}
{Caleb}, M., {Flynn}, C., {Bailes}, M., {Barr}, E.~D., {Hunstead}, R.~W.,
  {Keane}, E.~F., {Ravi}, V.,  \& {van Straten}, W. 2015, ArXiv e-prints

\bibitem[\protect\citeauthoryear{{Champion} et~al.}{{Champion}
  et~al.}{2015}]{champ2015}
{Champion}, D.~J., et~al. 2015, ArXiv e-prints

\bibitem[\protect\citeauthoryear{Close et~al.}{Close et~al.}{2010}]{close2010}
Close, S., Colestock, P., Cox, L., Kelley, M.,  \& Lee, N. 2010, Journal of
  Geophysical Research: Space Physics, 115, n/a, A12328

\bibitem[\protect\citeauthoryear{{Cordes} et~al.}{{Cordes}
  et~al.}{2006}]{cordes2006}
{Cordes}, J.~M., et~al. 2006, \apj, 637, 446

\bibitem[\protect\citeauthoryear{{Danish Khan}}{{Danish
  Khan}}{2014}]{danish2014}
{Danish Khan}, M. 2014, ArXiv e-prints

\bibitem[\protect\citeauthoryear{{Deng} \& {Zhang}}{{Deng} \&
  {Zhang}}{2014}]{deng2014}
{Deng}, W.,  \& {Zhang}, B. 2014, \apjl, 783, L35

\bibitem[\protect\citeauthoryear{{Dodin} \& {Fisch}}{{Dodin} \&
  {Fisch}}{2014}]{dolin2014}
{Dodin}, I.~Y.,  \& {Fisch}, N.~J. 2014, \apj, 794, 98

\bibitem[\protect\citeauthoryear{{Dolag} et~al.}{{Dolag}
  et~al.}{2015}]{dolag2015}
{Dolag}, K., {Gaensler}, B.~M., {Beck}, A.~M.,  \& {Beck}, M.~C. 2015, \mnras,
  451, 4277

\bibitem[\protect\citeauthoryear{{Katz}}{{Katz}}{2014}]{katz_perytons}
{Katz}, J.~I. 2014, \apj, 788, 34

\bibitem[\protect\citeauthoryear{{Keane} et~al.}{{Keane}
  et~al.}{2016}]{keane2016}
{Keane}, E.~F., et~al. 2016, \nat, 530, 453

\bibitem[\protect\citeauthoryear{{Keane} \& {Petroff}}{{Keane} \&
  {Petroff}}{2015}]{keane2015}
{Keane}, E.~F.,  \& {Petroff}, E. 2015, \mnras, 447, 2852

\bibitem[\protect\citeauthoryear{{Kulkarni} et~al.}{{Kulkarni}
  et~al.}{2014}]{kulkarni2014}
{Kulkarni}, S.~R., {Ofek}, E.~O., {Neill}, J.~D., {Zheng}, Z.,  \& {Juric}, M.
  2014, \apj, 797, 70

\bibitem[\protect\citeauthoryear{{Law} et~al.}{{Law} et~al.}{2015}]{law2015}
{Law}, C.~J., et~al. 2015, \apj, 807, 16

\bibitem[\protect\citeauthoryear{{Li} et~al.}{{Li} et~al.}{2016}]{li2016}
{Li}, L., {Huang}, Y., {Zhang}, Z., {Li}, D.,  \& {Li}, B. 2016, ArXiv e-prints

\bibitem[\protect\citeauthoryear{{Lorimer} et~al.}{{Lorimer}
  et~al.}{2007}]{lorimer2007}
{Lorimer}, D.~R., {Bailes}, M., {McLaughlin}, M.~A., {Narkevic}, D.~J.,  \&
  {Crawford}, F. 2007, Science, 318, 777

\bibitem[\protect\citeauthoryear{{Macquart} \& {Johnston}}{{Macquart} \&
  {Johnston}}{2015}]{macquart2015}
{Macquart}, J.-P.,  \& {Johnston}, S. 2015, \mnras, 451, 3278

\bibitem[\protect\citeauthoryear{{Masui} et~al.}{{Masui}
  et~al.}{2015}]{masui2015a}
{Masui}, K., et~al. 2015, ArXiv e-prints

\bibitem[\protect\citeauthoryear{{Masui} \& {Sigurdson}}{{Masui} \&
  {Sigurdson}}{2015}]{masui2015b}
{Masui}, K.~W.,  \& {Sigurdson}, K. 2015, Physical Review Letters, 115, 121301

\bibitem[\protect\citeauthoryear{{McQuinn}}{{McQuinn}}{2014}]{mcquinn2014}
{McQuinn}, M. 2014, \apjl, 780, L33

\bibitem[\protect\citeauthoryear{{Oppermann}, {Connor}, \& {Pen}}{{Oppermann}
  et~al.}{2016}]{oppermann2016}
{Oppermann}, N., {Connor}, L.,  \& {Pen}, U.-L. 2016, ArXiv e-prints

\bibitem[\protect\citeauthoryear{{Petroff} et~al.}{{Petroff}
  et~al.}{2015}]{petroff2015}
{Petroff}, E., et~al. 2015, \mnras, 447, 246

\bibitem[\protect\citeauthoryear{{Petroff} et~al.}{{Petroff}
  et~al.}{2016}]{frbcat}
{Petroff}, E., et~al. 2016, ArXiv e-prints

\bibitem[\protect\citeauthoryear{{Ravi}, {Shannon}, \& {Jameson}}{{Ravi}
  et~al.}{2015}]{ravi2015}
{Ravi}, V., {Shannon}, R.~M.,  \& {Jameson}, A. 2015, \apjl, 799, L5

\bibitem[\protect\citeauthoryear{{Saint-Hilaire}, {Benz}, \&
  {Monstein}}{{Saint-Hilaire} et~al.}{2014}]{saint2014}
{Saint-Hilaire}, P., {Benz}, A.~O.,  \& {Monstein}, C. 2014, \apj, 795, 19

\bibitem[\protect\citeauthoryear{{Siemion} et~al.}{{Siemion}
  et~al.}{2012}]{siemion2012}
{Siemion}, A.~P.~V., et~al. 2012, \apj, 744, 109

\bibitem[\protect\citeauthoryear{{Spitler} et~al.}{{Spitler}
  et~al.}{2014}]{spitler2014}
{Spitler}, L.~G., et~al. 2014, \apj, 790, 101

\bibitem[\protect\citeauthoryear{{Spitler} et~al.}{{Spitler}
  et~al.}{2016}]{spitler2016}
{Spitler}, L.~G., et~al. 2016, \nat, 531, 202

\bibitem[\protect\citeauthoryear{{Staveley-Smith} et~al.}{{Staveley-Smith}
  et~al.}{1996}]{smith1996}
{Staveley-Smith}, L., et~al. 1996, PASA, 13, 243

\bibitem[\protect\citeauthoryear{{Thornton} et~al.}{{Thornton}
  et~al.}{2013}]{thornton2013}
{Thornton}, D., et~al. 2013, Science, 341, 53

\bibitem[\protect\citeauthoryear{{Zheng} et~al.}{{Zheng}
  et~al.}{2014}]{zheng2014}
{Zheng}, Z., {Ofek}, E.~O., {Kulkarni}, S.~R., {Neill}, J.~D.,  \& {Juric}, M.
  2014, \apj, 797, 71

\end{thebibliography}

\appendix
\section{A1. Multiple-beam detection statistics}
\begin{figure*}
\centering
\includegraphics[width=\linewidth]{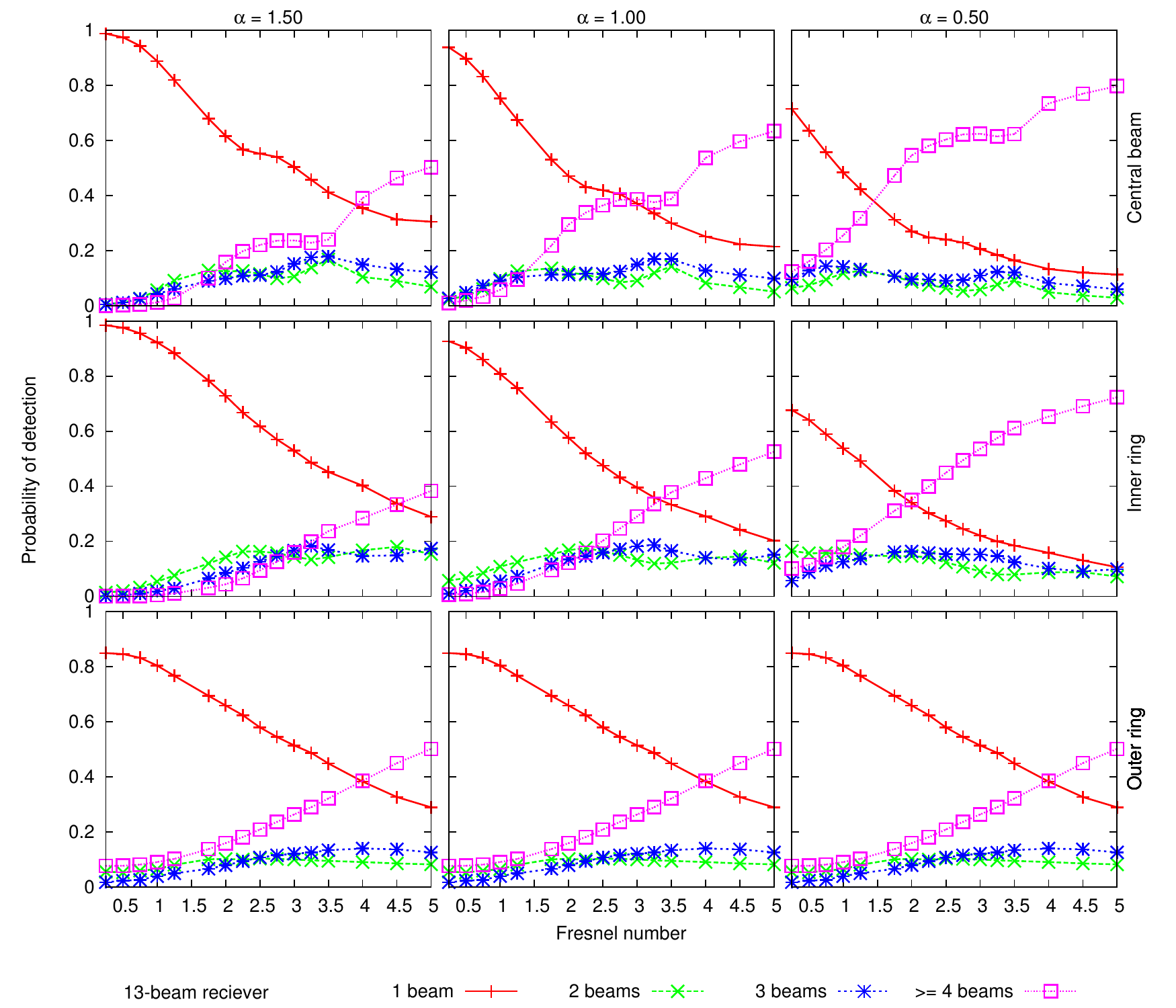}
\caption{Probabilities of multiple-beam detections at $\nu=1.5$~GHz for the Parked 13-beam receiver as functions of Fresnel number $n_{\rm f}=\frac{4d^2}{\lambda D}$ ($d=$antenna diameter, $D=$ source distance) of the FRB source. The panel columns correspond to varying logarithmic slopes of the integral source, and the three rows correspond to the cases where the central beam and an inner-ring beam are the principal beam. \label{fig:prob13}}
\end{figure*}

The probability of $n$-beam detections depends on the geometry of the feed-horn arrangement on the focal plane, the size and focal ratio of the dish, wavelength, and relative detection thresholds of the feeds. We assume a focal ratio of 0.41 for the Parkes dish.  Using the procedure described in Section~\ref{sec:pmbe}, we have computed the probabilities of a burst being detection in $i$ beams ($i=1,2,3...13$) for different principal beams, as a function of the Fresnel number of the source. Fig.~\ref{fig:prob13} shows a plot of these probabilities. The three rows correspond to cases where the central, an inner-ring and an outer-ring feed horn forms the principal beam. The columns correspond to values of the log$N$-log$F$ parameter, $\alpha$, of 1.5, 1.0, and 0.5. $\alpha=1.5$ corresponds to the case of non-evolving sources in Euclidean space and lower values of $\alpha$ give progressively flatter log$N$-log$F$ curves i.e an increasing propensity for brighter events. As the Fresnel number increases, the source moves closer to the telescope and appears progressively defocused at the focal plane, as a result of which, is detected in multiple beams with increasing probability. The probability for multiple-beam detection increases when the central feed forms the principal beams, since the central feed has more neighbors than the inner and outer ring feeds. All principal beams lead to very low probabilities ($\sim 1$ in 100) of multiple beam detection for $\alpha=1.5$ (see also Table~\ref{tab:mbp}), for far field sources ($n_{\rm f}\lesssim 1$). \\

The relative detection thresholds for the Parkes 13-beam feed array for the central, inner ring and outer ring feeds are taken into account by scaling the simulated beam-gains with $\eta^{-1}_{\rm eff}T^{-1}_{\rm rec}$, where $T_{\rm rec}$ is the receiver temperature.  The quoted values for $\eta_{\rm eff}$ are 1.36, 1.45, and 1.72~Jy~K$^{-1}$ respectively. The quoted values for $T_{\rm rec}$ are 29, 30, and 36~K respectively\footnote{\url{http://www.atnf.csiro.au/research/multibeam/.overview.html}}. The multiple-beam probabilities are, however, only affected by the ratio of detection thresholds between different feeds.
\section{A2. Multi-telescope detection statistics}

The various FRB surveys used to constrain the log$N$-log$F$ slope (see Table~2) are described here.

\begin{itemize}
\item \emph{ASSERT}: The ASSERT program observed with two antennas: a log-periodic dipole, and a horn antenna. \citet{saint2014} quote an SNR of 10 for a 2.5 K event lasting for 10~ms. Based on this and their bandwidth of $560$~MHz, we use a system temperature of 850~K. Since ASSERT does not use dish antennas, we have assumed an aperture efficiency of $\eta_{\rm eff}=1$. Given the large FWHM of the dipole (110~$\times$~70~deg$^2$), we have assumed a equivalent dish diameter of $\lambda/2=0.1$~m akin to a dipole antenna. For the horn antenna to obtain a FWHM of 10~deg we have assumed $d=1.2$~m. We expect these approximations to affect the conversion from antenna temperature to flux-density at few tens of percent level. Given the inability of current experiments (save the VLA) to obtain an accurate localization and hence an accurate flux-density, these approximations are justified.\\

\item \emph{ATA}: The Allen Telescope Array parameters are somewhat difficult to incorporate in our unified analysis since \citet{siemion2012} observed with 14 of the 30 antenna in dual-pol mode and the rest in single-pol mode. \citet{siemion2012} quote a single-pol SEFD of 10~kJy as an average for the 44 single-pol inputs used in the analysis. This corresponds to a single-pol system temperature of 120~K, which we used in our analysis. In addition, since the ATA observations were in Fly's Eye mode, i.e each antenna was pointed to a different sky location, take $N_{\rm ant}=1$ and multiply the total observing time with 30 which is the number of independent concurrent pointings. \citet{siemion2012} used 580 input~$\cdot$~days of data which corresponds to $N_{\rm day}$ = 580 / (44 inputs) $\times$ (30 antenna) / (24~hr/day) = 16.47 days.\\

\item \emph{Arecibo}: \citet{spitler2014} quote values of 10.4 and 8.2~K~Jy$^{-1}$ for the central and inner-ring beams of the Arecibo multiple-beam receiver. We take a weighted average of 8.5~K~Jy$^{-1}$, which corresponds to a $d=220$~m aperture with an efficiency of $\eta_{\rm eff}=60$\%. The Arecibo receivers have a $T_{\rm sys}$ of $30$~K.\\

\item \emph{VLA}: The VLA is a special case of a search for FRBs in interferometric images. Since the search was limited to the FWHM of the VLA dishes, for the VLA case, we have restricted the angular integration in Equation~\ref{eqn:ndet} to the FWHM. In addition, during interferometric imaging, the signals from the $N_{\rm ant}=27$ VLA dishes were combined coherently, and thus $N_{\rm ant}$ was replaced with $N_{\rm ant}^2$ in Equation~\ref{eqn:th_noise} for the VLA. 
 
\end{itemize}

\section{A3. Population statistics in Euclidean space}
Let the intrinsic burst energy and its observed fluence be $\fint$ and $\fobs$ respectively. Let a non-evolving population of FRB sources be distributed between distances of $\rmin$ and $\rmax$, and let the number of sources per unit volume with intrinsic energies between $\mathcal{F}$ and $\mathcal{F}+{\rm d}\mathcal{F}$ be $\rho(\mathcal{F})$. Then, $\fint$ and $\fobs$ for a source at distance $R$ are related by
\begin{equation}
\fobs = \frac{\fint} {4\pi R^2}
\end{equation} 
Sources with intrinsic energy $\fint$ will be observed to have a fluence in excess of $\fint$ if they are within a distance of $\left[\fint/(4\pi\fobs) \right]^{0.5}$. The total number of sources with intrinsic energy $\fint$ that have an observed fluence larger than some value $\fobs$ is then given by
\begin{eqnarray}
\frac{ {\rm d} \mathcal{N}(>\fobs,\fint)}{ {\rm d} \fint} &=& 0 \,\,\,\,\,\,\, \fint<4\pi\rmin^2\fobs  \nonumber\\
&=&\rho(\fint) \int_{\rmin}^{\sqrt{\frac{\fint}{4\pi\fobs}}} {\rm d}R\, 4\pi R^2; \,\,\,\,\,\,\, 4\pi\rmin^2 \fobs < \fint < 4\pi\rmax^2\fobs \nonumber \\
 &=& \rho(\fint) \int_{\rmin}^{\rmax} {\rm d}R\, 4\pi R^2; \,\,\,\,\,\,\, \fint > 4\pi \rmax^2 \fobs \\
\end{eqnarray}

Evaluating the integrals, and then integrating over $\fint$ gives
\begin{eqnarray}
\label{eqn:nfobs}
\mathcal{N}(>\fobs) &=& \frac{4\pi}{3}\int_{4\pi\fobs \rmin^2}^{4\pi\fobs\rmax^2} {\rm d}\fint \rho(\fint) \left[\fint^{1.5} (4\pi\fobs)^{-1.5}-\rmin^{3} \right]\nonumber \\
&& + \frac{4\pi\left( \rmax^3-\rmin^3\right)}{3} \int_{4\pi\fobs\rmax^2}^{\infty}{\rm d}\fint \rho(\fint) 
\end{eqnarray}
Assuming $\rho(\fint)=\propto \fint^{-\beta-1}$, which yields an intrinsic energy distribution with a log$N$-log$F$ index of $\beta$, the integrals can be evaluated analytically:
\begin{equation}
\label{eqn:nfobs_red}
\mathcal{N}(>\fobs) \propto \frac{\fobs^{-\beta}}{\beta(1.5-\beta)} \left( \rmax^{3-2\beta}-\rmin^{3-2\beta} \right).
\end{equation}
We have shown that in Euclidean space, in the presence of a minimum and/or maximum distance to the population, the log$N$-log$F$ parameter for the observed fluences is the same as that of the intrinsic energy distribution. Furthermore, the number of detected events within a sphere of radius $\rmax$ scales as $\rmax^{3-2\beta}$, or the number of events from a infinitesimally thin shell of thickness ${\rm d}R$ at radius $R$ scales as $R^{2(1-\beta)}$. For $\beta<1$, the detected population is biased towards larger distances, and for $\beta=1$ there is no distance bias in the detected population. Note that we have implicitly assumed that $\fint^{\rm min}<4\pi\fobs\rmin^2$, and $\fint^{\rm max}>4\pi\fobs\rmax^2$. The former is a reasonable assumption, but the latter will break down for very large values of $\rmax$, at which point, $\mathcal{N}(>\fobs)$ will saturate (for $\beta<1.5$)  and equation \ref{eqn:nfobs_red} will not longer be valid.\\

We can treat the `standard candle' scenario as follows. In the absence of any distance evolution in $\rho(\fint)$, the observed fluence distribution can be obtained from Equation~\ref{eqn:nfobs} by substituting $\rho(\fint) = \delta(\fint-\mathcal{L}_0)$, where $\mathcal{L}_0$ is the standard-candle energy, and $\delta(.)$ is the Dirac delta function. We assume that $\mathcal{L}_0$ is finite and set $R_{\rm max}$  to some high value such that $\mathcal{L}_0<\fobs R^2_{\rm max}$. Under these conditions, the second integral in Equation~\ref{eqn:nfobs} goes to zero, and the first integral yields the observed fluence distribution under the standard candle hypothesis:
\begin{equation}
\mathcal{N}(>\fobs) = \frac{4\pi}{3} \left[\mathcal{L}_0^{1.5} \left(4\pi\fobs\right)^{-1.5} -R_{\rm min}^3 \right]\,\,\,\,\, {\rm (standard~candle)}
\end{equation}
For small values of $R_{\rm min}$, the index of the log$N$-log$F$ function is $\alpha=1.5$, as expected. If $\mathcal{L}_0>\fobs R^2_{\rm max}$, then the first integral in Equation~\ref{eqn:nfobs} reduces to 0, and the second integral yields $\mathcal{L}_0$. The observed log$N$-log$F$ function becomes independent of $\fobs$ i.e $\alpha=0$ which is strongly disfavored by our constraints.\\

Motivated by our findings that strongly disfavor $\alpha=1.5$, we have considered a `toy model' where FRBs are standard candles and $\rho(\fint)$ evolves with distance as $\rho(R) \propto R^{\kappa}$. In this case, the integrations over $R$ and $\fint$ are coupled, but for the standard-candle case, the algebra is greatly simplified. All events within a distance of $\left[\fint/(4\pi\fobs)\right]^{0.5}$ will have an observed fluence in excess of $\fobs$. Hence, the observed fluence distribution may  be evaluated as:
\begin{equation}
\mathcal{N}(>\fobs) = \int_{0}^{\sqrt{\frac{\fint}{4\pi\fobs}}} \rho(R) 4\pi R^2 {\rm d}R \propto \frac{4\pi}{3} \left(\frac{\fint}{4\pi\fobs} \right)^{\frac{k+3}{2}}
\end{equation}
Hence the relationship between the log$N$-log$F$ parameter $\alpha$ and the distance evolution parameter $\kappa$ is $\kappa = 2\alpha-3$. The bounds on $\kappa$ corresponding to the 90\% bounds $0.66<\alpha<0.96$ are $-1.68<\kappa<-1.08$ at 90\% confidence. We find such a distance-evolution law to be a contrived arrangement since physical parameters that may contribute to FRB rates such as galaxy counts and star-formation rate do not adhere to such laws. Based on this, the standard-candle hypothesis is strongly disfavored.

\section{A4. Statistics for a cosmological population}
For a cosmological population, we can follow the same steps as that for a local population with the inclusion of the effects of (i) redshift evolution of comoving volume element and luminosity distance, and (ii) effects of time dilation on the fluence due to cosmic expansion. Fluence has units of erg~m$^{-2}$~s, which unlike flux-density which has units of erg~m$^{-2}$, is affected by time dilation. We will express all distances in units of the Hubble distance. $\fint$ and $\fobs$ are then related as
\begin{equation}
\fobs = \frac{\fint (1+z)}{(1+z)^2r^2(z)}
\end{equation}
where the denominator is the square of the luminosity distance, $(1+z)$ in the numerator accounts for time-dilation due to cosmic expansion, and $r(z)$ is the radial coordinate which is in-turn given by
\begin{eqnarray}
r(z) &=& \int_{0}^{z} {\rm d}z'E(z') \nonumber \\
E(z) &=& \sqrt{\Omega_{\rm m}(1+z)^3 + \Omega_{\Lambda}}
\end{eqnarray}
The number of detected events above some threshold fluence $\fobs$ is then given by
\begin{eqnarray}
\mathcal{N}(>\fobs) &=& \int_{0}^{\infty}{\rm d}\fint \rho(\fint)\,\,\int^{\Psi^{-1}\left( \frac{\fint}{\fobs}\right)}_{0} {\rm d}V(z) \nonumber\\
{\rm d}V(z) &=& 4\pi\frac{r^2(z)}{E(z)}{\rm d}z
\end{eqnarray}
where $\Psi(z) = (1+z)r^2(z)$, is the ratio between the intrinsic energy and observed fluence, and ${\rm d}V(z)$ is the comoving volume element.\\

In this formalism, the effects of color-corrections and intrinsic source evolution can be incorporated easily. If an burst has a spectral index $\gamma$, that is $\fint(\nu)\propto \nu^{-\gamma}$, then we have a modified relationship between $\fint$ and $\fobs$: $\Psi(z,\gamma) = (1+z)^{1-\gamma}r^2(z)$. Similarly, any function of redshift that describes the evolution of intrinsic source counts may be taken into the redshift integral over the comoving volume element.\\

Finally, the cumulative number of events from sources out to some redshift $z_{\rm max}$ can be evaluated as
\begin{equation}
\mathcal{N}(>\fobs,<z_{\rm max}) =\int_{0}^{\fobs\Psi(z_{\rm max})}{\rm d}\fint \rho(\fint)\,\,\int^{\Psi^{-1}\left( \frac{\fint}{\fobs}\right)}_{0} {\rm d}V(z) + \int_{\fobs\Psi(z_{\rm max})}^{\infty}{\rm d}\fint \rho(\fint)\,\,\int_{0}^{z_{\rm max}} {\rm d}V(z)
\end{equation}
The above integrals must again be computed numerically. We assume the following cosmological parameters: $\Omega_{\rm m}=0.25$, and $\Omega_{\rm \Lambda}=0.75$. 
\end{document}